\documentstyle[twocolumn,aps,epsfig]{revtex}
\begin{document}
\draft
\title{Higgs Decay into Gravitons: Trees and Loops}
\author{A. Widom}
\address{Physics Department, Northeastern University, Boston MA U.S.A.}
\author{Y.N. Srivastava }
\address{Physics Department \& INFN, University of Perugia, Perugia, Italy}
%%%%%%%%%%%%%%%%%%%%%%%%%%%%%%%%%%%%%%%%%%%%%%%%%%%%%%%%%%%%%%%%%%%%%%%%
\maketitle
%%%%%%%%%%%%%%%%%%%%%%%%%%%%%%%%%%%%%%%%%%%%%%%%%%%%%%%%%%%%%%%%%%%%%%%%
\begin{abstract}
The decay of the Higgs particle into two gravitons was 
previously calculated by us using a Born term from the 
Einstein field equations. Subsequently, others computed 
the same decay via one loop diagrams but omitting the 
Born terms. Here all of the diagrams up to one loop are 
discussed, and it is shown that the Born term is overwhelmingly 
dominant in agreement with our original results.
\end{abstract}
\vskip .4cm

From the viewpoint of the standard electroweak model, all of the 
inertial mass is due to the Higgs particles so it is quite evident 
that almost all of experimental gravity must be the result of the 
Higgs field; i.e. the inertial and gravitational mass of an object 
are the same. With this physics in mind we calculated\cite{1} 
the decay of the Higgs into two gravitons 
\begin{equation}
H\to g+g.
\end{equation}
The action for the decay was computed from the trace of the stress 
tensor
\begin{equation}
T=g^{\mu \nu }T_{\mu \nu }
\end{equation}
via the action 
\begin{equation}
S_{eff}=\left({1\over c\left<\phi \right>}\right)\int \chi T d\Omega
\end{equation}
where the Higgs scalar field is written 
\begin{math}\phi =\left<\phi \right>+\chi \end{math} and 
the space time volume element 
\begin{math} d\Omega =\sqrt{-g}d^4x \end{math}. To obtain the Born term, 
shown below in Fig.1, we employed the Einstein field equations for 
the stress tensor
\begin{equation}
T_{\mu \nu }=\left({c^4\over 8\pi G}\right)
\left(R_{\mu \nu }-{1\over 2}g_{\mu \nu}R\right).
\end{equation}

\begin{figure}[htbp]
\begin{center}
\mbox{\epsfig{file=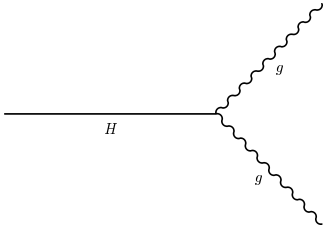,height=30mm}}
\caption{Shown is the Higgs to two graviton Born term.}
\label{hfig1}
\end{center}
\end{figure}

Subsequent to our work, the decay in Eq.(1) was computed via 
the quantum corrections\cite{2} to the stress. A typical one loop 
diagram which was considered is shown below in Fig.2. The one loop order  
correction to the stress tensor (the gravitational Casimir effect) 
has the form\cite{3} 
$$
T^{(1)}_{\mu \nu}=\left({\hbar c \over 2880 \pi^2}\right)\times 
$$
\begin{equation}
\left(
a_1 R^{\alpha \beta }R_{\alpha \mu \beta \nu }+ 
a_2 R^2 g_{\mu \nu }+a_3{\delta \over \delta g^{\mu \nu}}
\int R^2 d\Omega 
\right),
\end{equation}
where the coefficients 
\begin{math}a_1,a_2 \end{math} and \begin{math} a_3 \end{math} are 
dimensionless and of order unity.

\begin{figure}[htbp]
\begin{center}
\mbox{\epsfig{file=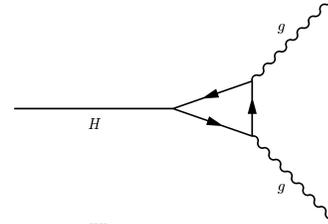,height=30mm}}
\caption{Shown is a Higgs to two graviton decay contribution 
which proceeds via an internal Fermion loop.}
\label{hfig2}
\end{center}
\end{figure}

The full expression for \begin{math} T  \end{math}, including the 
Einstein (classical tree diagrams) and the Casimir (one loop)  
correction terms may be written 
$$
T_{total}=-\left({c^4 R\over 8\pi G}\right)+
\left({\hbar c \over 2880 \pi^2}\right)\times 
$$
\begin{equation}
\left(
b_1 R^2 +b_2 R^{\mu \nu }R_{\mu \nu }
+b_3R^{\mu \nu \alpha \beta }R_{\mu \nu \alpha \beta }
+b_4 \Box R
\right)+ ...\ .
\end{equation}
where \begin{math} \{b_j\} \end{math}  
(\begin{math} j=1,2,3,4 \end{math}) are (again) constants of order 
unity.

In our previous work we employed the first term, which contains 
the Newtonian coupling strength \begin{math} G  \end{math} but not 
Planck's constant \begin{math}   \end{math}. Our work includes 
the stress tensor term which contains almost all of usually 
discussed gravity effects which have been observed, or at least  
may soon be observed\cite{4}. The stress tensor term we included 
contains the Newtonian gravitational law, the perihelion anomaly 
in the orbit of Mercury, gravitational waves and the bending of 
light around the Sun. 

The Einstein term, which contains the Newtonian coupling strength 
was thrown away\cite{2} by the workers who computed Eq.(1) employing 
{\em only the second term} in Eq.(6). This amounts to treating 
very precisely the very tiny terms and throwing away the very big 
terms. For example, if the Higgs-gravity interaction were really of the 
strength of the gravitational Casimir effect, then the Higgs induced 
inertial mass would never induce a sufficient gravitational strength 
to explain Kepler's laws of planetary orbits. The Higgs mechanism 
would already be ruled out if only one loop gravity terms were to be 
considered. 

To see in detail from where the large terms arise, one may examine 
the gravitational action 
\begin{equation}
S_{g}=\left({c^3\over 16\pi G}\right)\int R d\Omega, 
\end{equation}  
and write \begin{math} R \end{math} as\cite{5,6} 
\begin{equation}
R=\left({1\over \sqrt{-g}}\right)
\partial_\mu \left(\sqrt{-g}W^\mu \right) + \tilde{R},
\end{equation}
where \begin{math} \tilde{R} \end{math} and 
\begin{math} W^\mu  \end{math} depend only on 
\begin{math} g_{\mu \nu } \end{math} and 
\begin{math} \Gamma^\mu _{\lambda \nu } \end{math}, i.e. 
on the metric and its first derivatives. The decomposition 
in Eq.(8) depends on the gauge (i.e. coordinate representation) for 
\begin{math} g_{\mu \nu } \end{math}, but the sum is gauge 
invariant. For tree diagrams, \begin{math} \hbar \to 0 \end{math} 
in Eq.(6), one finds from Eqs.(3) and (8) 
\begin{equation}
S_{eff}=\left({c^3\over 8\pi G\left<\phi \right>}\right)
\int \left(W^\mu \partial_\mu \chi -\tilde{R}\chi \right) d\Omega .
\end{equation} 
Eq.(9) generates all of the {\em tree diagrams} for the Higgs to decay 
into an arbitrary number \begin{math} N \end{math} gravitons. 
The tree level coupling depends only on the fields and their first 
derivatives. This is essential in any field theory. Each 
emitted graviton carries a factor 
\begin{math} \propto \sqrt{G} \end{math} 
as can be seen in Eq.(13) below.

For small deviations from flat space time, 
\begin{equation}
g_{\mu \nu }=\eta_{\mu \nu }+h_{\mu \nu },
\end{equation}
and in the gravitational Lorentz gauge, the 
\begin{math} H\to g+g  \end{math} tree diagram of Fig.1 is generated 
from Eq.(9) using the second order term 
\begin{equation}
R^{(2)}=\left({1\over 4}\right)
\left(\partial_\lambda h^{\mu \nu }\partial^\lambda h_{\mu \nu }\right)
-\left({1\over 8}\right)
\left(\partial_\lambda h \partial^\lambda h\right),
\end{equation}
where \begin{math} h=\eta^{\mu \nu }h_{\mu \nu }\end{math}. Thus, one finds 
the action for the tree diagram in Fig.1 
\begin{equation}
S(H\to g+g)=-\left({c^3\over 8\pi G\left<\phi \right>}\right)
\int R^{(2)} \chi d^4 x. 
\end{equation}

In operator form, one sees that the action in Eqs.(11) and (12) 
can create or destroy  two gravitons by inserting the second 
quantized field 
$$
h_{\mu \nu }({\bf r},t)=\sqrt{32\pi \hbar G\over c^3}
\sum_{\lambda=\pm 2}\int 
\left({d^3 {\bf k}\over (2\pi )^3 2|{\bf k}|}\right)\times
$$
\begin{equation}
\left\{
a({\bf k},\lambda )\epsilon_{\mu \nu }({\bf k},\lambda )
e^{i({\bf k\cdot r}-\omega t)}+
a^\dagger ({\bf k},\lambda )\epsilon^*_{\mu \nu }({\bf k},\lambda )
e^{-i({\bf k\cdot r}-\omega t)}
\right\} 
\end{equation}
where \begin{math} \omega =c|{\bf k}| \end{math}, the physical 
helicity polarizations obey 
\begin{math}   
\epsilon_{\mu \nu }({\bf k},\lambda )
\epsilon^{\mu \nu *}({\bf k},\lambda^\prime  )=
\delta_{\lambda ,\lambda^\prime },
\end{math}
and 
\begin{equation}
\left[
a({\bf k},\lambda ),a^\dagger ({\bf k}^\prime ,\lambda^\prime  )
\right]
=(2\pi )^3 (2|{\bf k}|)\delta^{(3)}({\bf k}-{\bf k}^\prime )
\delta_{\lambda \lambda^\prime }.
\end{equation}
It is {\em very remarkable} that the Newtonian coupling strength 
cancels out when all of the Eqs.(11), (12) and (13) are taken 
into account. The tree diagram of Fig.1 yields a transition 
rate {\em independent} of \begin{math} G \end{math}.

The reader may wish to check that the Feynman quadratic action\cite{7}    
\begin{equation}
S^{(2)}[h]=\left({c^3\over 16\pi G}\right)\int R^{(2)}d^4x,
\end{equation}
when employed in the path integral for the stress action 
\begin{math} W[T] \end{math}, 
\begin{equation}
e^{(iW[T]/\hbar )}=\int \ e^{(iS_g[h]/\hbar )}
\ e^{(i\int T^{\mu \nu }h_{\mu \nu}d^4x /2\hbar c)}
\ {\cal D}h,
\end{equation}
leads to the Schwinger result 
$$
W[T]=
$$ 
\begin{equation}
{1\over 2c^5}\int d^4x \int d^4y 
T^{\mu \nu }(x)D_{\mu \nu \lambda \sigma }(x-y)T^{\lambda  \sigma  }(y).
\end{equation}
The graviton propagator is given by 
$$
D_{\mu \nu \lambda \sigma }(x-y)=
$$ 
\begin{equation}
\left(\eta_{\mu \lambda }\eta_{\nu \sigma }+
\eta_{\mu \sigma }\eta_{\nu \lambda }-
\eta_{\mu \nu }\eta_{\lambda \sigma }\right)D(x-y),
\end{equation}
where Newton's gravitational law is made manifest by 
\begin{equation}
D(x)=-\int \left({4\pi G \over Q^2-i0^+}\right)e^{iQ\cdot x}
\left({d^4Q\over (2\pi )^4}\right).
\end{equation}
Schwinger has argued\cite{4} that Eq.(17), which is based on Eq.(11), 
describes all of experimental Newtonian gravity and most of 
experimental general relativity. This is as reliably locally 
Lorentz invariant as is presently possible.

Finally, the tree diagram of Fig.1 leads to the Higgs decay rate 
\begin{equation}
\Gamma (H\to g+g)=\left({\sqrt{2}\over 16\pi }\right)
\left({G_F M_H^2\over \hbar c}\right)\left({M_Hc^2\over \hbar }\right),
\end{equation}
where \begin{math} G_F \end{math} 
is the Fermi coupling strength, 
\begin{math} (\sqrt{2} \left<\phi \right>^2)=c^3/(\hbar G_F) \end{math}, 
and \begin{math} M_H  \end{math} is the Higgs particle. 
We stand by our previously reported result.

\end{document}